\RequirePackage{fix-cm}
\documentclass[smallextended]{svjour3}       
\smartqed  
\usepackage{graphicx}
\usepackage{mathbbol}
\usepackage{esvect}
\usepackage{amsmath}
\usepackage{pdflscape}
\usepackage{amssymb}
\usepackage{endnotes}
\usepackage{mathbbol}
\usepackage{amstext}
\usepackage{natbib}
		\setcitestyle{authoryear,round,aysep={}}
\begin{document}

\title{Modeling Managerial Search Behavior based on Simon's Concept of Satisficing
}


\author{Friederike Wall}


\institute{Friederike Wall \at
              University of Klagenfurt, Department for Management Control and Strategic Management, Universit\"atsstrasse 65-67, 9200 Klagenfurt, Austria \\
              Tel.: +43-463-27004031\\
              \email{friederike.wall@aau.at}   
}

\date{Received: date / Accepted: date}

\maketitle

\begin{abstract}
Computational models of managerial search often build on back\-ward-looking search based on hill-climbing algorithms. Regardless of its prevalence, there is some evidence that this family of algorithms does not universally represent managers’ search behavior. Against this background, the paper proposes an alternative algorithm that captures key elements of Simon’s concept of satisficing which received considerable support in behavioral experiments. The paper contrasts the satisficing-based algorithm to two variants of hill-climbing search in an agent-based model of a simple decision-making organization. The model builds on the framework of NK fitness landscapes which allows controlling for the complexity of the decision problem to be solved. The results suggest that the model’s behavior may remarkably differ depending on whether satisficing or hill-climbing serves as an algorithmic representation for decision-makers’ search. Moreover, with the satisficing algorithm, results indicate oscillating aspiration levels, even to the negative, and intense -- and potentially destabilizing -- search activities when intra-organizational complexity increases. Findings may shed some new light on prior computational models of decision-making in organizations and point to avenues for future research.
\end{abstract}

\section{Introduction}
\label{intro}

Computational models of managerial search often comprise adaptive processes based on experiential learning and backward-looking search behavior (\citealp[e.g.,][]{Gavetti2000,Kollman2000,Dosi2003,Ethiraj2004,Siggelkow2005,Wall2017}).
In computational models of managerial search, for capturing experiential learning and backward-looking search behavior, hill-climbing algorithms prevail (\citealp[for overviews see][]{Ganco2009,Baumann2019}). Based on local search, a particular feature of these algorithms is that a decision-maker would \textit{never} accept or preserve performance-decreasing changes (\citealp[e.g.,][]{Altenberg1997,Russell2016}). This feature of hill-climbing algorithms has been criticized regarding cognitive biases such as escalation of commitment, overconfidence, and confirmation bias (\citealp[e.g.,][]{Staw1981,Astebro2014,Mercier2017}). \citet{Tracy2017} recently question that hill-climbing algorithms are appropriate representations of managerial search behavior. Based on experimental findings, they suggest studying alternative algorithmic representations of managerial search and their effects on model behavior with respect to findings of prior research.  

The research presented here follows this line of argumentation. The paper seeks to contribute to computational management science by proposing and exemplarily applying an alternative algorithm for representing managerial search behavior. In particular, the paper introduces an algorithm for experiential learning and backward-looking search for managers based on Herbert A. Simon's concept of satisficing\footnote{The term satisficing results from a merger of the two words: \textit{satis}fying and suf\textit{ficing} in the sense that in the process a solution is found which is both satisfying and sufficing \citep{Hoffrage2004}.} \citep{Simon1955} which has turned out being a relevant representation of search behavior \citep[e.g.,][]{Gueth2010,Caplin2011}. According to Simon, satisficing means searching sequentially for options until the decision-maker regards the level of utility achieved as satisfactory. The aspiration level shapes what is regarded as satisfactory. The aspiration level and the maximum number of options searched -- depending on the difficulty of the decision problem to be solved -- may be subject to adaptation.  

Against this background, the paper has a twofold research objective: 
\begin{enumerate}
	\item introduction of an algorithm for managerial search behavior according to Simon's satisficing concept;
	\item exemplary application of the satisficing algorithm in contrast to hill-climbing algorithms in an agent-based simulation to figure out potential differences and commonalities regarding model behavior.
\end{enumerate}

For this, the paper proceeds as follows: The next section provides an overview of the theoretical background with particular focus on Simon's idea of satisficing, before, in Section \ref{satisficing}, the algorithm capturing core elements of satisficing is introduced in the context of searching for superior solutions of combinatorial decision problems. 

Next, the proposed satisficing algorithm is contrasted to hill-climbing algorithms. This is done via the example of an agent-based simulation model of organizations operating on rugged performance landscapes. The performance landscapes are modeled according to the NK framework as initially introduced in evolutionary biology \citep{Kauffman1987, Kauffman1993}. The rationale for this choice is that many models dealing with search in organizations build on the NK framework \citep[for overviews, e.g.,][]{Baumann2019,Wall2016}. Hence, the NK model serves as a kind of ``quasi-standard'' in research on managerial search. This makes the NK model a functional basis for the second research objective mentioned above. A particular feature of the NK model is that it allows to systematically vary the complexity of a search problem in terms of the interdependencies among its sub-problems \citep{Li2006,Csaszar2018} which makes it appropriate to study the search behavior for varying levels of difficulty to locate the global maximum. Hence, the illustrative agent-based simulation model presented controls for the level of intra-organizational complexity among subordinate managerial decision-makers. This is particularly relevant in view of the satisficing concept since the difficulty of finding satisfactory solutions drives adjustments, for example, of the aspiration level. The model is outlined in Section 4.

Section 5 introduces the experimental settings for the simulations. The simulations are conducted for purposes of explanation and prediction \citep{Za2018,Burton2011} with particular focus on the differences that satisficing vs. hill-climbing search entail for the model behavior. The results are presented and discussed in Section 6 followed by  concluding remarks.

\section{Search and saticficing: Foundations and related work }
\subsection{Preliminary remarks on the theoretical background}
In traditional schools of economic thinking, economic actors know, at least in principle, the entire space of solutions for their decision problems. Knowing the whole search space allows them to behave as utility maximizers, i.e., detecting and choosing that option out of the solution space which maximizes the respective utility function \citep{vonNeumann2007}. \citet{Simon1955} claimed that there is an ``absence of evidence that the classical concepts describe the decision-making process'' (p. 104). Among Simon's arguments is that information gathering on options and their outcomes may not be costless. 

However, the cost of search and information has been introduced taking a ``classical economic perspective''. \citet{Stigler1961} claimed that information on options often is not known in advance but has to be searched, and this may reasonably bring about search costs. 
Accordingly, in making the concept of utility maximizing more realistic, a decision-maker has to solve a sophisticated problem of economic choice: whether or not, to incur the search cost for better information which requires to forecast the information's benefits (i.e., better choices) in terms of all its future consequences including subsequent choices. Yet, it has been argued that this extended ``version'' of the utility maximizing model, though economically stringent, does not capture real situations of decision-making for several reasons -- among them principal problems of mathematical tractability or cognitive limitations \citep[e.g.,][]{Conlisk1996,Gigerenzer2002,Gigerenzer2004}. \citet{Gigerenzer2002} argues that the rule to stop searching for information when the cost exceeds benefits \citep{Stigler1961} may paradoxically require more time, knowledge, and computational abilities of decision-makers (``sophisticated econometricians'') than in models with unbounded rationality.

\subsection{Search in computational management science and its algorithmic representation}

Against this background, a large body of research in computational management science, particularly in the vein of agent-based computational economics \citep{Tesfatsion2003,Chang2006,Chen2012}, is based on the concept of bounded rationality \citep{Simon1955,Simon1959}. In particular, it is often assumed that economic agents do not dispose of a ``theoretical'' understanding of their problems, including knowledge of the solution space \citep[an exception is ][]{Gavetti2000}; instead, agents have to search stepwise for superior solutions, e.g., solutions that provide better outcome with respect to the objective than the status quo \citep{Safarzynska2010}. Hence, in computational models of search, instead of global optimization -- with or without constraints imposed by the cost of information -- agents often conduct experiental learning and backward-looking search. This is represented by local search, meaning that only one or some attributes of the current state (or policy) are changed; should this change be productive compared to the status quo, the modified policy serves as basis for a new local search. This results in adaptive processes. However, there is evidence that adaptive processes based on experiential learning are biased against new alternatives (\citealp[e.g.,][]{Levinthal1981,Levinthal1997}), especially since adaptation does not correct early sampling errors (hot-stove effect) \citep{Denrell2001}. 

With the shift from ``instantaneous'' global optimization to stepwise and local search also the processual perspective -- including questions of speed of performance enhancements and of contingent factors -- comes into play. In particular, the complexity of decision problems and environmental turbulence are among the predominant contingent factors in the respective stream of research. Computational studies on search behavior have been carried out in various domains like, for example, organizational design, innovation, psychology, and, accordingly, the related approaches in prior research are rather manifold. Overviews are, for example, given in \citet{Ganco2009}, \citet{Wall2016}, or \citet{Baumann2019}.



In computational studies capturing backward-looking search behavior, greedy algorithms and, in particular, hill-climbing algorithms predominate. According to \citet[][p. 414]{Cormen2009}, a ``greedy algorithm always makes the choice that looks best at the moment'' in terms of ``a locally optimal choice in the hope that this choice will lead to a globally optimal solution''. A hill-climbing algorithm -- employing the metaphor of seeking the highest summit -- for a move in the landscape requires that the outcome (``altitude'') will increase. In other words: the aspiration level is a performance improvement of greater than zero. 
For example, with a steepest ascent hill-climbing algorithm, that option out of more than one alternatives to the status quo is selected which provides the highest improvement in outcome; if none of the alternatives promises an incline in outcome, the status quo is kept. With this, hill-climbing algorithms are particularly prone to get stuck in local maxima, ridges, or plateaus in a landscape \citep[for overviews, e.g.,][]{Cormen2009,Macken1991,Selman2006}. This is mainly because with these algorithms a short-term decline in favor of a long-term increase would not happen since no choice in favor of an option that provides an inferior outcome than the status quo would ever be made. Hence, hill-climbing algorithms may lead to rather myopic search processes. Moreover, as mentioned in the Introduction, it was argued that this is also be in conflict with some cognitive biases which indicate that decision-makers eventually behave in favor of performance declines. These considerations gave rise to questions whether hill-climbing algorithms appropriately capture managerial search behavior \citep[e.g.,][]{Tracy2017}.


While hill-climbing algorithms are customary in computational studies capturing managerial search processes, it is worth mentioning that they often serve just as the nucleus: in many models, managerial search is embedded in a broader context. This context is, for example, defined by the incentive schemes shaping managers' objective functions and, thus, the particular ``landscapes'' managers are searching in \cite[e.g.,][]{Siggelkow2005,Wall2017}. Another contingency factor is the imprecision of managers' information, which may, accidentally, lead to short-term declines but long-term inclines of performance choices \citep{Knudsen2007,Wall2016}. Furthermore, prior research studied the decomposition of the organizational decision problem \citep{Dosi2003} or the coordination among managers searching on partitions of the overall decision problem \citep{Siggelkow2003}. Moreover, the learning-based adaptation of the \textit{structure} of search comes into play. For example, based on experience, the organization of search processes (e.g., who searches on which particular decision problem) could be subject to coevolution \citep[e.g.,][]{Wall2018b}.



However, as mentioned before, the potential of hill-climbing algorithms to represent managerial search behavior has been questioned, and the very core of the research endeavor presented in this paper is to introduce and illustrate an algorithm for backward-looking search based on Simon's concept of satisficing.
 
\subsection{Outline of Simon's concept of satisficing} \label{Simon}

This section intends to provide an overview of the ``satisficing'' concept with particular focus on an algorithmic representation for backward-looking search.\footnote{An in-depth analysis of the concept's origin in Simon's work is given by \citet{Brown2004}; \citet{Radner1975} introduced a general formulation for purposes of mathematical optimization.} 
The following quote captures the core idea  \citep[][p. 110]{Simon1955}:  

\begin{footnotesize}
\begin{quote}
\noindent	``In most global models of rational choice, all alternatives are evaluated before a choice is made. In actual human decision-making, alternatives are often examined sequentially. We may, or may not, know the mechanism that determines the order of procedure. When alternatives are examined sequentially, we may regard the first satisfactory alternative that is evaluated as such as the one actually selected.''
\end{quote}
\end{footnotesize}

\noindent The satisficing concept is explained and justified extensively in Simon's 1955 paper and subsequent works (\citealp[e.g.,][]{Simon1959,Simon1979}; \citealp[for a reconstruction of satisficing from Simon's early works see][]{Brown2004}). One of Simon's arguments is that decision-makers endowed with limited information-processing capabilities may strive for decisions which are good enough with reasonable costs of computation (\citealp[][pp. 106]{Simon1955}; \citealp[][p. 498]{Simon1979}). The quote above indicates on three building blocks which are particularly relevant for an algorithmic representation of satificing. These are:\footnote{These three elements, in principle, correspond to building blocks proposed by \citet{Gigerenzer1999} for heuristics, which are search rules, stopping rules and decision rules \citep[see also][]{Gigerenzer2011}.}  
\begin{enumerate}
	\item \textit{sequential procedure}, i.e., options are discovered \textit{and} evaluated sequentially; 
	\item \textit{aspiration level}, i.e., options are evaluated with respect to a level of outcome that is regarded satisfactory; 
	\item \textit{stopping rule}, i.e., search is stopped when the first satisfactory option is found.
\end{enumerate}

\noindent Regarding the stopping rule, Simon introduces further considerations to assure, first, that -- at least in the long run -- a satisfactory alternative can be found while, second, in the short-term, search can provisionally stop if no satisfactory alternative is identified. For this, in particular, he introduces a dynamic perspective by considering a \textit{sequence of situations} with choices to be made \citep[][p. 111]{Simon1955}: 

\begin{footnotesize}
\begin{quotation}
\noindent ``The aspiration level, which defines a satisfactory alternative, may change from point to point in this sequence of trials. A vague principle would be that as the individual, in his exploration of alternatives, finds it \textit{easy} to discover satisfactory alternatives, his aspiration level rises; as he finds it \textit{difficult} to discover satisfactory alternatives, his aspiration level falls. Perhaps it would be possible to express the ease or difficulty of exploration in terms of the cost of obtaining better information about the mapping of A on S, or the combinatorial magnitude of the task of refining this mapping. There are a number of ways in which this process could be defined formally.
Such changes in aspiration level would tend to bring about a `near-uniqueness' of the satisfactory solutions and would also tend to guarantee the existence of satisfactory solutions. For the failure
to discover a solution would depress the aspiration level and bring satisfactory solutions into existence.'' [emphasis in original]
\end{quotation}
\end{footnotesize}

\noindent As Simon points out, such a mechanism of adjusting aspiration levels assures that, satisfactory solutions exist in the long run. However, as mentioned before, a second aspect is the number of alternatives a decision-maker is willing to explore. In short-term such an upper bound assures that the search, in principle, may stop even if no satisfactory option is found; however, in a sequence of situations, the maximum number of alternatives searched may be subject to adjustment too \citep[][p. 111]{Simon1955}:  

\begin{footnotesize}
\begin{quotation}
\noindent ``Up to this point little use has been made of the distinction between $A$, the set of behavior alternatives, and $\dot{A}$, the set of behavior alternatives that the organism considers. Suppose now that the latter is a proper subset of the former. Then, the failure to find a satisfactory alternative in $\dot{A}$ may lead to a search for additional alternatives in $A$ that can be adjoined to $\dot{A}$.'' 
\end{quotation}
\end{footnotesize}

\noindent Simon mentions these two types of adjustment -- i.e., regarding aspiration levels and maximum number of options searched -- as examples of how decision-making behavior could be adjusted to the perceived difficulty of finding satisfactory alternatives. Moreover, the two types of adjustments may substitute or complement each other \citep[][p. 112]{Simon1955}: 

\begin{footnotesize}
\begin{quotation}
\noindent ``In one organism, dynamic adjustment over a sequence of choices may depend primarily upon adjustments of the aspiration level. In another organism, the adjustments may be primarily in the set $\dot{A}$: if satisfactory alternatives are discovered easily, $\dot{A}$ narrows; if it becomes difficult to find satisfactory alternatives, $\dot{A}$ broadens... The more persistent the organism, the greater the role played by the adjustment of $\dot{A}$, relative to the role played by the adjustment of the aspiration level.''
\end{quotation}
\end{footnotesize}

\noindent Hence, for an algorithmic representation, the above ``list'' of building blocks of satisficing could be extended by

\begin{enumerate}
	\item[4.] \textit{adjustment of aspiration level}, with down\-ward (up\-ward) ad\-just\-ment when the de\-ci\-sion-ma\-ker finds it difficult (easy) to identify a satisfactory alternative; 
	\item[5.] \textit{adjustment of maximum number of options explored}, with broadening (narrowing) adjustment when the decision-maker finds it difficult (easy) to identify a satisfactory alternative.
\end{enumerate}

The concept of satisficing stimulated a large body of further research in various domains reaching from psychology and economics to multi-agent systems \citep[e.g.,][]{Bianchi1990,Gigerenzer2002,Todd2003,Parker2007,Schwartz2008,Rosenfeld2012}. For example, key elements of satisficing are among the foundations of ``the adpative toolbox'' comprising ``fast and frugal heuristics'' introduced by \citet{Gigerenzer2002}. Moreover, Simon's satisficing provides a basis for Selten's prominent ``aspiration adaption theory'' \citep{Selten1998,Selten2002}. However, particularly the idea of aspiration levels has given rise to questions on how they are initially set and how they are updated \citep[e.g.,][]{Bianchi1990,Gueth2007,Schwartz2008}. A further body of research seeks to test how far satisficing captures real human decision-making behavior empirically. For example, in an experimental study \citet{Caplin2011} find considerable support for key elements of satisficing behavior, namely sequential search and stopping a search process when a decision-maker regards the level of outcome satisfactory. Another stream of research refers to the conditions when decision-makers seek to behave as maximizers or satisficers, i.e., to styles of decision-making \citep[e.g.,][]{Schwartz2002,Parker2007}. 

\section{Algorithmic representation of satisficing managerial search behavior}\label{satisficing}

\subsection {Preliminary remarks}
This section introduces a computational model of organizations with decision-making agents that employ satisficing in backward-looking search behavior following Simon's concept as introduced in Section \ref{Simon}. The model is presented for decision-making agents facing a multidimensional binary decision problem. This modeling choice builds on two arguments: 

First and most important, as it was outlined above, a considerable body of research in computational management science employs the prominent NK framework. In its standard form, the NK framework comprises $N$-dimensional binary bit strings as the vector of choices or features adapted throughout adaptive processes based on some kinds of learning or evolution. Hence, modeling the satisficing concept for binary decision problems eases the integration into prior research.\footnote{In Section 4, this paper presents simulation experiments for managerial decision-makers within organizations. The organizations face a binary decision problem according to the NK-framework. The subsequent description of satisficing search behavior follows some notational conventions of the NK framework.} 

Second, a fixed dimensionality binary decision problem facilitates to model satisficing search behavior. For example, the maximum number of alternatives (see Section \ref{Simon}) and the term neighborhood can be figured out easily. However, the author believes that the simplifying assumption of binary decision problems does not limit, in principle, transferring the proposed algorithm of satisficing search behavior to other types of decision problems.

\subsection{Process structure of satisficing search}\label{structure}

Subsequently, satisficing search behavior of a manager $r$ is described where manager $r$ may be one out of $M$ managers in an organization (i.e., $r=(1,\ldots,M)$). Manager $r$ faces an $N^r$-dimensional binary decision problem. 

According to the behavioral assumptions of \citet{Simon1955}, manager $r$ is not able to survey the \textit{entire} search space and, hence, cannot ``locate'' the optimal solution of its decision problem ``at once''. Instead, manager $r$ employs a time-consuming search process to identify solutions with superior performance, or even the optimal solution, regarding manager $r$'s objective.

As outlined in Section \ref{Simon}, a particular feature of satisficing search behavior is that, when searching for superior performance, an agent may adapt the aspiration level and the maximum number of alternatives discovered before the agent decides to stop searching. Hence, the proposed model comprises three adaptive processes which are related to each other: In each period $t$ of time,
\begin{enumerate}
	\item manager $r$ sequentially searches for novel options to its particular decision problem within the institutional framework given which includes, for example, division of labor or rewards provided (Section \ref{sequence});
  \item manager $r$ adjusts the aspiration level $a^r$ that a newly found option will have to meet to be selected in the next period based on the performance improvements resulting from the solutions implemented in the past (Section \ref{aspiration});
	\item manager $r$ adjusts the maximum number $s^{max,r}$ of options to be discovered before search is stopped depending on the number of options that manager $r$ had to search for before a satisficing alternative was found in the past (Section \ref{searchmax}).
\end{enumerate}

\begin{figure}[!ht]
\centering
  \includegraphics[width=1\textwidth]{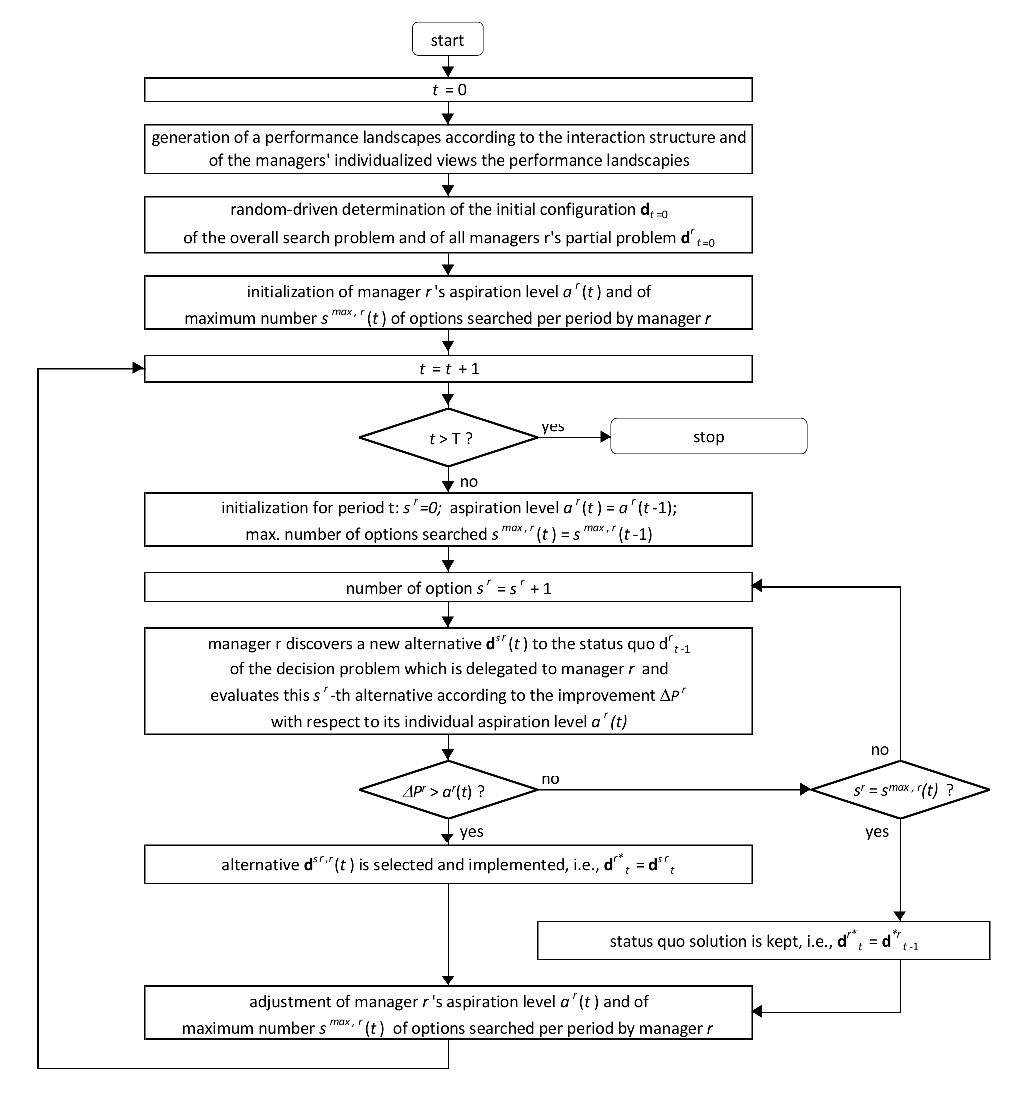}
\caption[ ]{Process structure of satisficing search behavior} \label{flowchart}
\end{figure}

\noindent Figure \ref{flowchart} shows the principle process of satisficing search behavior of a manager $r$. Subsequently, the model is described in more detail.

\subsection{Sequential Search for New Options}\label{sequence}
A key feature of satisficing search is that new options are discovered \emph{and} evaluated sequentially: the agent discovers one novel option $\vec{d}^{s^r}_t$ and evaluates (i.e., searches for ``cues'' in the terminology of \citet{Simon1955} whether it promises a performance improvement compared to the status quo $\vec{d}^{r}_{t-1}$ that, at least, meets the aspiration level $a^r(t)$, i.e., when 
\begin{equation}
  \Delta P^r_t \geq a^r(t)
\end{equation}
with 
\begin{equation}
	\Delta P^r_t = P(\vec{d}^{s^r}_t) - P(\vec{d}^{r}_{t-1})\label{perfdiff}
\end{equation}

If so, this option is implemented, and search is stopped for this time step $t$; otherwise, the next option is searched and evaluated as far as the maximum number of options $s^{max,r}(t)$ is not reached yet (see Figure \ref{flowchart}).  

With manager $r$ facing an $N^r$-dimensional binary decision problem, at maximum, $2^{N^r}-1$ alternative configurations $\vec{d}^{r}$ compared to the status quo could be implemented. Hence, the upper bound for he maximum number of options is 
\begin{equation}
	 s^{max,r} \leq 2^{N^r}-1 
\end{equation}

For an algorithmic representation of satisficing, defining a sequence of the agent's discoveries of new options is necessary. For the sequence of options' discovery, various possibilities could are feasible. For example, one obvious way is to let the agent \textit{randomly} discover one out of the $2^{N^r}-1$ alternatives (if an option has been discovered before in that time step $t$, the random draw is repeated). 

However, the simulation experiments presented subsequently employ a \textit{``closest-first'' search policy} which reflects the idea of neighborhood search: a manager $r$ starts searching in the immediate ``neighborhood'' of the status quo. Should this not lead to a satisficing option, manager $r$ extends the ``circle'' of search around the status quo. Hence, the sequence follows increasing Hamming distances of discovered alternatives to the status quo where the Hamming distances of an alternative option $\vec{d}^{s^r}_t$ to the status quo is given by
\begin{equation}
   h(\vec{d}^{s^r}_t) = \sum^{N^r}_{i=1} {\left|\vec{d}^{r}_{t-1} - \vec{d}^{s^r}_t\right|}\label{hamming}
\end{equation}

Hence, the search starts with alternatives with a Hamming distance $h(\vec{d}^{s^r}_t)=1$, then followed by options with a Hamming distance of two and so forth, as long as neither the aspiration level is met nor the maximum number of options $s^{max,r}$ to be considered is reached. Among the options with equal Hamming distance the sequence is given at random.\footnote{For example, with a decision problem of $N^r=3$, three alternatives to the status quo with a Hamming distance $h(\vec{d}^{s^r}_t)=1$, three alternatives with $h(\vec{d}^{s^r}_t)=2$ and one with $h(\vec{d}^{s^r}_t)=3$ exist. A manager first discovers nearest neighbors; next, options with $h(\vec{d}^{s^r}_t)=2$ are found etc. where the sequence among equal-distanced options is randomly given.\label{altspace}}

The rationale for a sequence given by increasing Hamming distances is as follows: This sequence appears particularly appropriate to capture the idea of stepwise improvement of a given configuration. With respect to the cost of search and change, small steps (i.e. Hamming distance equal to 1) could be assumed to show lower cost than more distant options which require more changes. Hence, the \textit{``closest-first'' search policy} may be based on considerations of cost of search and change. 

However, it is worth mentioning that other forms of the sequence of searching are arguable too: For example, a manager may be rewarded based on the particular novelty of the options chosen, which could give reason to start searching with the most distant alternatives possible.

\subsection{Adaptation of the aspiration level} \label{aspiration}

As mentioned in Sections \ref{Simon} and \ref{structure}, a core element in satisficing is the aspiration level. Newly found options are evaluated according to whether or not they promise to meet the aspiration level, and the aspiration level is subject to adaptation based on experience \citep{Simon1955}: The aspiration level may increase (decrease) depending on how easy (difficult) it was to find a satisfactory alternative in the past.

In the proposed model of satisficing search behavior, the aspiration level is adjusted according to the performance experience, i.e., an improvement or deterioration of performance (see Eq. \ref{perfdiff}) achieved over time. In particular, the aspiration level $a^r(t)$ is captured as an exponentially weighted moving average of past performance changes where $\alpha^r$ denotes the speed of adjustment for manager $r$ \citep{Levinthal1981,Boergers2000,Levinthal2016}, i.e.,

\begin{equation}
%
 a^r(t+1) = \alpha^r\cdot \Delta{P^r}_{t} +(1-\alpha^r)\cdot a^r(t). \label{aspirationadjust}
\end{equation}  

It is worth emphasizing that the aspiration level could also become negative -- i.e., a performance \textit{decline} becoming acceptable -- if declines happened in the past. This establishes a contrast to hill-climbing algorithms where decision-makers would not accept performance declines (see Introduction). Section \ref{baseline} comes back to this aspect.

\subsection{Adaptation of the maximum number of options searched} \label{searchmax}

In a similar vein, the space of options in which a manager searches for satisfactory alternatives may be dynamically adjusted. When it turns out to be difficult to find satisfactory options, the search space for alternatives is broadened; when finding satisfactory options is easy, search space is narrowed \citep{Simon1955}.

In the modeling effort presented here, this is captured as adjustment of the maximum number $s^{max,r}$ of options that the decision-making agent $r$ may consider in the next time step. In particular, if in period $t$ a maximum number of options, i.e.,  $s^{r}(t)=s^{max,r}$, was searched and evaluated without that a satisfactory alternative to the status quo was identified, then for $t+1$ the (potential) search space increases. For this, again, an exponentially weighted moving average of past search spaces is employed where $\beta^r$ denotes the speed of adjustment for manager $r$. Hence, the search space results from

\begin{equation}
s^{max,r}(t+1) 
=\left\{\begin{array}{ll}
          \beta^r\cdot (s^{r}(t)+1) +(1-\beta^r)\cdot s^{max,r}(t) \\  
					    \mathrm{\hphantom{000}if\hphantom{0}} s^{r}(t)=s^{max,r}(t) 
						   \mathrm{\hphantom{0}and\hphantom{0}} \Delta{P^r}_{t} < a^r(t)\\
							\\
	       \beta^r\cdot (s^{r}(t)) +(1-\beta^r)\cdot s^{max,r}(t) \\  
	                        \mathrm{\hphantom{000}else}			
				\end{array}\right.	\label{spaceadjust}								
\end{equation}    

However, since the maximum search space $s^{max,r}$ has to be an integer, the moving average according to the upper case of Eq. \ref{spaceadjust} is to be rounded up or down which is done according to
\begin{equation}
   s^{max,r}(t+1) = \left\lfloor s^{max,r}(t+1) + 0.5 \right\rfloor \label{spaceround}
\end{equation}

Hence, with Eq. \ref{spaceround}, the ``adjusting'' procedure in Eq. \ref{spaceadjust} does not necessarily result in an adjusted space $s^{max,r}(t+1)$ of options for the next period.

\section{Example: An agent-based model of search in collaborative organizations based on satisficing vs. hill-clim\-bing}

\subsection{Overview}
The model of satisficing agents as outlined in the preceding section is studied for artificial organizations that seek superior solutions of binary decision problems according to the NK-framework \citep{Kauffman1987, Kauffman1993}. A particular purpose is to contrast the adaptive walks of organizations resided by satisficing managers to organizations with managers employing a hill-climbing algorithm as familiar in the domain of agent-based computational organization science. 

First, the overall organizational decision problem, its decomposition, and delegation to managers are introduced (Section \ref{decproblem}). Next, a description of managers' objective functions and information basis (Section \ref{objectives}) follows. Third, search and decision-making via hill-climbing are briefly outlined in contrast to satisficing (Section \ref{hillclimbing}).


\subsection{Decision Problem and Structure of the Organizations} \label{decproblem}

In the simulation model, artificial organizations are observed while searching for superior solutions for a decision problem according to the framework of NK-fitness landscapes. In particular, at each time step $t$ the organizations face an $N$-dimensional binary decision problem, i.e., $\vec{d_t}={(d_{1t},...,d_{Nt})}$ with $d_{it}\in \left\{0,1\right\}$, $i=1,...,N$, out of $2^{N}$ different binary vectors possible. Each of the two states $d_{it}\in \left\{0,1\right\}$ provides a distinct contribution $C_{it}$ to the overall performance $V(\vec{d_t})$. The contributions $C_{it}$ are randomly drawn from a uniform distribution with $0\leq{C_{it}}\leq1$. Parameter $K$ (with $0\leq K\leq N-1$) reflects the number of those choices $d_{jt}$, $j\neq{i}$ which also affect the performance contribution $C_{it}$ of choice $d_{it}$. Hence, $K$ captures the complexity of the decision problem in terms of the interactions among decisions: this means that contribution $C_{it}$ may not only depend on the single choice $d_{it}$ (being 0 or 1) but also on $K$ other choices: 
\begin{equation}
C_{it}=f_{i}(d_{it};d_{i_{1}t},...d_{i_{K}t}),
\end{equation}
with $\left\{i_1,...,i_K\right\} \subset\left\{1,...,i-1,i+1,...,N\right\}$. In case of no interactions among choices, $K$ equals 0, and $K$ is $N-1$ for the maximum level of complexity where each single choice $i$ affects the performance contribution of each other binary choice $j\neq i$. 
The overall performance $V_t$ achieved in period $t$ results as normalized sum of contributions $C_{it}$ from
\begin{equation}
	V_{t}=V(\vec{d_t})=\frac{1}{N}\sum^{N}_{i=1} {C_{it}}\label{perf}.
\end{equation}

The organizations have a hierarchical structure and comprise two types of agents: (1) one headquarter and (2) $M$ managers. The organizations make use of division of labor. In particular, the $N$-dimensional overall decision problem is decomposed into $M$ disjoint partial problems, and each of these sub-problems is exclusively delegated to one manager $r=(1,\ldots,M)$. For the sake of simplicity, the sub-problems are of equal size $N^r$.\footnote{With $N \in \mathbb{N}$ this requires that $N$ is divisible by $M$ without remainder.} Each manager $r$ is endowed with decision-making authority on its ``own'' partition of the organization's decision problem. 

The headquarter seeks to maximize the overall performance $V_{t}$ according to Eq. \ref{perf}. However, its role is restricted to -- at the end of each time step $t$ -- observing the overall performance $V_{t}$, observing each manager’s performance contribution and rewarding managers accordingly.

Depending on the complexity $K$ of the $N$-dimensional decision problem and the particular structure of interactions among the $M$ sub-problems, indirect interactions among the managers' choices may result. Let $K^{ex}$ denote the level of interdependencies across managers' sub-problems. In case that interdependencies across sub-problems exist, i.e., if $K^{ex}>0$, then the performance contribution of manager $r$'s choices to overall performance $V$ is affected by choices made by other managers $q \neq r$ and vice versa (see, for example, Figure \ref{interactions}.b).

\subsection{Managers' Objective Functions and Information} \label{objectives}

The managers seek to maximize compensation which is merit-based and depends on the performance contribution $P^{r}_{t}(\vec{d_t})$ of manager $r$ to overall performance $V(\vec{d_t})$ according to Eq. \ref{perf}. Hence, we have

\begin{equation}
	V_{t}=V(\vec{d_t})=\sum^{M}_{r=1} {P^{r}_{t}(\vec{d_t})}\label{perf2} 
\end{equation}
with 
\begin{equation}
	 P^{r}_{t}(\vec{d^r_t})=\frac{1}{N}\sum^{N^r}_{i=1+w} {C_{it}} \label{perfmanager}
\end{equation}
\noindent and with ${w}={\sum^{r-1}_{m=1} {N^m}}$ for $r>1$ and ${w}=0$ for ${r}=1$. 

For the sake of simplicity, compensation of manager $r$ depends linearly on the value base $P^{r}_{t}(\vec{d_t})$ for all levels of $P^{r}_{t}$. Hence, by increasing the performance contribution $P^{r}_{t}$ of the partial solution for the $N^r$-dimensional sub-problem to the overall organization's decision-problem, manager $r$ also increases its compensation.  

However, when making their choices on their respective partial configurations $\vec{d^r_t}$, the managers show some further cognitive limitations (apart from not knowing the entire space of solutions and, thus, having to search for options): 

First, manager $r$ cannot anticipate the other departments' $q\neq{r}$ choices; rather manager $r$ assumes that the fellow managers will stay with the status quo, i.e., opt for $\vec{d}^{q*}_{t-1}$. 

Second, manager $r$ is not able to perfectly ex-ante evaluate the effects of any newly discovered option $\vec{d}^{s^r}_t$ on the value base for compensation $P^{r}_{t}(\vec{s^d}^{r}_t)$
(see Eq. \ref{perfmanager}). Rather, ex ante evaluations are afflicted with noise which is, for the sake of simplicity, an relative error imputed to the actual performance (\citealp{Wall2010}; for further types of errors see \citealp{Levitan1995}; for further models of  managerial search capturing imperfect evaluations see \citealp{Carley1997,Chang1998,Knudsen2007}). The error terms ${e}^{r}(\vec{d}^{s^r}_t)$ follow a Gaussian distribution $N(0;\sigma)$ with expected value $0$ and standard deviations $\sigma^{r}$; errors are assumed to be independent from each other. Hence, the value base of compensation $\tilde{P}^{r}_{t}(\vec{d}^{s^r}_t)$ of a newly discovered $\vec{d}^{s^r}$ option as ex ante \emph{perceived} by manager $r$ is 

\begin{equation}
 {\tilde{P}^{r}_{t}(\vec{d}^{s^r}_t)={P}^{r}_{t}(\vec{d}^{s^r}_t)+{e}^{r}(\vec{d}^{s^r}_t)} \label{discompsum}
\end{equation}

\noindent Thereby, when making decisions, each manager $r$ has a different ``view'' of the actual fitness landscape which results from (1) the decomposition of the overall decision problem and the delegation of sub-problems and (2) from the managers' individual ``perceptions'' due to the individualized error terms $\sigma^{r}$. 
However, for the status quo option $\vec{d}^{r*}_{t-1}$, it is assumed that manager $r$ remembers the compensation from the last period. From this, manager $r$ also knows the \textit{actual} performance ${P}^{r}_{t}$ of the status quo, should the manager choose to stay with it in time step $t$ and if, in case of interactions across sub-problems, also the fellow managers stay with the status quo.

\subsection {Search strategies} \label{hillclimbing}
\label{search}

In every time step $t$, each manager $r$ seeks to identify a superior configuration for its partial decision problem $\vec{d^{r}_{t}}$ with respect to the value base of compensation. The search strategy shapes the options a manager can choose.
The simulation model contrasts adaptive walks of organizations with \textit{satisficing} managers to those organizations with managers employing a \textit{steepest ascent hill-climbing algorithm} as frequently employed in computational management science. In Section \ref{satisficing}, the model of satisficing search was introduced. Hence, at this point, a short outline of hill-climbing in the context of the model follows. 

In particular, as already mentioned, in the model the managers cannot survey the \textit{entire} search space and, hence, they have to search stepwise for superior solutions. Following a hill-climbing algorithm, a manager searches in the neighborhood for a fixed number $s^{max,r}$ of alternatives and opts for an alternative only if it promises a higher performance (``fitness'') than the status quo. The distance to the status quo defines the term neighborhood and, in the context of the NK-model, is measured by the Hamming distance $h(\vec{d}^{s^r}_t)$ of an alternative option to the status quo $\vec{d}^{r}_{t-1}$ according to Eq. \ref{hamming}. 

In the most simple case, the neighborhood is set to $h(\vec{d}^{s^r}_t)=1$ and the number of alternatives is $s^r = s^{max,r}=1$, too. This means that only one alternative to $\vec{d}^{r}_{t-1}$ is discovered where -- usually at random -- one bit is flipped. However, the ``allowed'' neighborhood of search could be broader than one, and also the number of alternatives the manager identifies could be higher than one. Both is often employed in models of organizational search (\citealp[e.g.,][]{Siggelkow2005,Wall2017}; \citealp[for overviews ][]{Chang2006,Baumann2019}). If the number of alternatives $s^{max,r}$ identified providing a performance incline is higher than one, that option with the highest incline is selected (steepest ascent hill-climbing). Hence, three aspects of this hill-climbing algorithm (HCA) appear noteworthy in comparison to the satisficing algorithm (see Section \ref{satisficing}): 
\begin{enumerate}
	\item In the HCA, the number $s^r$ of newly discovered alternatives per period equals the maximum number of alternatives allowed, i.e., $s^r = s^{max,r}$. Moreover, the maximum number of alternatives is not subject to adaptation based on experience over time like in satisficing.\footnote{In this sense, the HCA could be regarded as a special case according to Eq. \ref{spaceadjust} with $\beta=0$.}
	\item The HCA employs an aspiration level of zero: alternatives with a performance incline compared to the status quo are worth being selected by a manager (i.e., $a^r > 0$). Additionally, unlike in satisficing, the aspiration level is not adapted according to experience.\footnote{With this, the HCA may be regarded as a special case according to Eq. \ref{aspirationadjust} with $\alpha=0$.}
	\item In the HCA, options are not searched \emph{and} evaluated in sequence with a stop of searching when an alternative meets the aspiration level like in satisficing. Instead, in case that the HCA is parametrized to two or more alternatives to be searched (i.e., if $s^{max,r}>1$), the search stops when $s^{max,r}$ alternatives are identified. Then these $s^{max,r}$ options are evaluated against the status quo and against each other to figure out the steepest ascent. 
\end{enumerate}


The paper presents the results of simulations for organizations with managers employing satisficing versus hill-climbing search. For this, the next section introduces the particular parameter settings of the simulation experiments.

\section{Simulation experiments and parameter settings}

The simulation study seeks to provide insights into how satisficing managerial search behavior compared to hill-climbing behavior affects the organizations’ resulting adaptive walks. Table \ref{Parameters} displays the parameter settings which are explained in the remainder of this section.

\begin{table} [ht]
\caption{Parameter settings}
\label{Parameters}
\begin{footnotesize}
\begin{center}
\begin{tabular}{ll} 
\hline\noalign{\smallskip}
Parameter & Values / Types \\
\noalign{\smallskip}\hline\noalign{\smallskip}
\multicolumn{2}{l}{\textit{Applying to all scenarios / types of managers}} \\
Observation period      & $T=250$ \\
Simulation runs         & per scenario 2,500 runs with 25 runs on 100 distinct \\ 
                        & fitness landscapes \\ \smallskip
Number of choices       & $N=12$ \\ 

Interaction structures  & decomposable: ($K=2; K^{ex}=0$) (see Fig. \ref{interactions}.a) \\ 
                        & non-decomposable: \\
												& - low: ($K=3; K^{ex}=1$); ($K=4; K^{ex}=2$); \\
												& - medium: ($K=5; K^{ex}=3$); ($K=6; K^{ex}=4$); \\ \smallskip
												& - high: ($K=7; K^{ex}=5$) (see Fig. \ref{interactions}.b)\\ 

Number of managers      & $M=4$ with $\vec{d}^{1}=(d_{1},d_{2},d_{3}), \vec{d}^{2}=(d_{4},d_{5},d_{6})$, \\ \smallskip
                        & $\vec{d}^{3}=(d_{7},d_{8},d_{9}),\vec{d}^{4}=d_{10},d_{11},d_{12})$\\
Managers' precision of  & $\sigma^{r}=0.05$ for all managers $r=(1,\ldots,M)$\\ \bigskip
ex-ante evaluation	\\										

\multicolumn{2}{l}{\textit{Satisficing type of managers}} \\
Aspiration level\\ 
\hphantom{0}- in the beginning     & $a^r(t=0)=0$ for all managers $r=(1,\ldots,M)$  \\ \smallskip
\hphantom{0}- speed of adjustment  & $\alpha^r = 0.5$ for all managers $r=(1,\ldots,M)$\\

Max. number of alternatives\\ 
\hphantom{0}- in the beginning     & $s^{max,r}(t=0)=2$ for all managers $r=(1,\ldots,M)$\\ \bigskip
\hphantom{0}- speed of adjustment  & $\beta^r = 0.5$ for all managers $r=(1,\ldots,M)$\\

\multicolumn{2}{l}{\textit{Hill-climbing type of managers}} \\

\textit{HC2}-strategy      & $s^{r}=2$ alternatives per period \\
									& with $h(\vec{d}^{r,a1})=1$ and $h(\vec{d}^{r,a2})=1$ \\ \smallskip
									& for all managers $r=(1,\ldots,M)$\\
									
\textit{HC6}-strategy  		& $s^{r} =6$ alternatives per period \\
									& with $h(\vec{d}^{r,a1})=1$, $h(\vec{d}^{r,a2})=1$, $h(\vec{d}^{r,a3})=1$,\\
									& and $h(\vec{d}^{r,a4})=2$, $h(\vec{d}^{r,a5})=2$, $h(\vec{d}^{r,a6})=2$ \\
									& for all managers $r=(1,\ldots,M)$\\
																																
\noalign{\smallskip}\hline
\end{tabular}

\end{center}
\end{footnotesize}
\end{table}

The parameter settings in the upper part of Table \ref{Parameters}) apply to experiments with both satisficing and hill-climbing types of managers. As such, organizations are observed for 250 periods\footnote{The observation period $T$ was fixed based on pretests which indicate that the results do not principally change for longer observation periods.} when searching for superior solutions to an $N=12$-dimensional decision problem. The overall decision problem is decomposed into $M=4$ equal-sized sub-problems of which each is delegated exclusively to a subordinate manager. 

\begin{figure}[!ht]
\centering
  \includegraphics[width=1\textwidth]{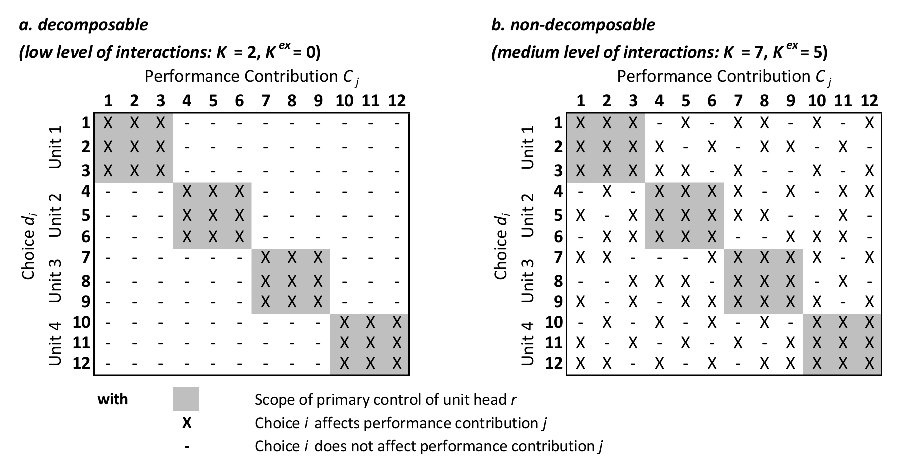}
\caption[ ]{Examples of decomposable and nearly decomposable interaction structures} \label{interactions} 
\end{figure}

The experiments are conducted for different levels of complexity of the organizations' decision problems: In particular, the organizations may have a perfectly decomposable interaction structure of decisions which captures situations where, for example, the task of an organization is perfectly decomposable along geographical regions or products without any interdependencies across regions or products, respectively \citep{Galbraith1974,Rivkin2007,Simon1962}. Figure \ref{interactions}.a gives an example of a situation with no interactions across managers' sub-problems (i.e., $K^{ex}=0$). Alternatively, the interaction structures captured in the experiments may exhibit a low, medium, or high level of interactions across sub-problems. For example, Figure \ref{interactions}.b shows a case of a high level of cross-problem interactions (i.e., $K^{ex}=5$). This interaction structure may represent situations caused by certain constraints of resources (budgets or capacities), by market interactions (prices of one product may affect the price of another) or functional interrelations (e.g., the product design sets requirements for procurement processes) \citep{Thompson1967,Galbraith1973,Rivkin2007}.  

When ex-ante evaluating newly discovered options, the managers suffer from some noise (see Eq. \ref{discompsum}) following a Gaussian distribution with mean 0 and a standard deviation of 0.05. This parametrization intends to reflect some empirical evidence according to which error levels around 10 percent could be a realistic estimation \citep{Tee2007,Redman1998}. 

Regarding experiments for organizations resided by satisficing managers (see middle part of Table \ref{Parameters}), the aspiration levels of performance enhancements start at a level of zero for two reasons: first, this corresponds to hill-climbing (see Section \ref{hillclimbing}) and, hence, eliminates one source of potential differences between the two modes in the experiments. Second, this ``conservative'' setting captures the desire to avoid, at least, situations of not-sustaining an already achieved performance level. For satisficing search, the maximum search space starts at a moderate level of just two alternatives, which also relates to a search space often specified for hill-climbing search in computational management science. Regarding the speed of adjustment for both the aspiration level of performance enhancements and the maximum number of alternatives, the present observation and the past are weighted equally with $\alpha^r$ (Eq. \ref{aspirationadjust}) and $\beta^r$ (Eq. \ref{spaceadjust}), respectively, set to 0.5.

The simulations experiments comprise two different steepest ascent hill-climbing strategies (see the lower part of Table \ref{Parameters}). In particular, in the ``HC2''-strategy, in every time step, each manager discovers two alternatives to the respective status quo, each alternative with one bit flipped compared to the status quo and thus captures local search. With the ``HC6''-strategy, in every time step, 6 alternatives to the current configuration are discovered, i.e., 3 with Hamming distance 1 and 3 with Hamming distance of 2.\footnote{Hence, in the HC6-strategy each manager $r$ identifies 6 out of the 7 possible alternatives to the $N^r=3$-dimensional partial decision problem, see fn. \ref{altspace}. The only option that is not feasible is switching each bit of the 3-dimensional sub-problem of each manager. The space of alternatives considered, could also be regarded as indication on managers' capabilities as in \citet{Rivkin2003}.} 

The HC2-strategy corresponds to agent-based models in prior research which study local search -- often in comparison to other forms of search (\citealp[e.g.,][]{Levinthal1997,Jain2014}) --  and, thus, serves as a basis for comparisons of simulation results obtained with satisficing agents. In contrast, the HC6-strategy serves another purpose in the experiments: it captures a kind of ``upper bound'' of feasible partial alterations given the overall decision-problem of the size $N=12$ and its decomposition into four equal-sized sub-problems. Hence, the HC6-strategy provides a broad search space and an obvious question is whether search spaces in satisficing (see Eq. \ref{spaceadjust}) may evolve to the same high level.

\section{Results and discussion}

In order to be clear and concise in exploring the parameter space, the results of the simulation experiments are presented in two steps. Following the idea of factorial design of simulation experiments \citep{Lorscheid2012}, Section \ref{baseline} introduces results of two baseline scenarios to analyze the principal effects of satisficing vs. hill-climbing managerial search behavior. In particular, organizations facing a decomposable decision-problem (i.e., $K^{ex}=0$) and organizations which have to deal with a medium level of complexity (i.e., $K^{ex}=3$) are studied. Section \ref{complexity} provides an analysis of the sensitivity to intra-organizational interactions for a broader range of complexity levels of the organization's decision-problem. 
  
\subsection{Baseline scenarios}\label{baseline}

Table \ref{baselineresults} reports condensed results obtained from the simulation experiments for the baseline scenarios. For each scenario (i.e., combination of interaction structure and search strategy), the respective 2500 simulation runs were analyzed with respect to several metrics.\footnote{In the analysis of simulation experiments, the metrics related to performance $V_t$ are given relative to the global maxima of the respective performance landscapes: otherwise, the results were not comparable across different performance landscapes.} 

The performance change achieved on average in the first ten periods informs about the speed of performance enhancement at the beginning of the adaptive walks, which may be particularly relevant in turbulent environments \citep{Siggelkow2005}. However, with respect to satisficing, the usually high performance inclines at the beginning of search are particularly interesting for the adjustments of aspiration levels and search spaces. The final performance, i.e., performance $V_{t=250}$ achieved in the last period of the observation time on average in the 2500 simulation runs per scenario, informs about the effectiveness of the search processes. This also applies to the relative frequency of how often the global maxima in the respective performance landscapes have been found in the 2500 simulation runs per scenario. The ratio of periods in which a new configuration $\vec{d_t}$ is implemented characterizes the adaptive walks more into detail.

\begin{table}[htbp]
\caption{Condensed results of baseline scenarios}
\label{baselineresults}
\begin{footnotesize}
\begin{center}
\begin{tabular}{lcccc}
\hline\noalign{\smallskip}
Search type	&Performance                &Final 	               &Frequ. of	   &Ratio of\\
                        &change in                      &performance  &glob. max. &periods\\
                        &first periods             &$V_{t=250}$   &found in     &with altered\\
                        &($V_{t=10}-V_{t=0}$)  &($\pm$CI*)     &$t=250$     &config. $\vec{d}$\\			\noalign{\smallskip}\hline\noalign{\smallskip}
\multicolumn{5}{l}{\textit{Decomposable interaction structure, $K^{ex}=0$ }}\\	
Satisficing	&+0.3110	&0.9928	$\pm$0.0010 &60.0\%	&22.4\% \\
HC2	hill-climbing   &+0.2609	&0.9507     $\pm$0.0031 &15.3\%	&12.3\% \\ \smallskip
HC6	hill-climbing   &+0.3207	&0.9960	$\pm$0.0007	&66.6\%	&20.6\% \\

\multicolumn{5}{l}{\textit{Non-decomposable interaction structure: medium, $K^{ex}=3$ }}\\	
Satisficing	&+0.1119	&0.8551	$\pm$0.0082	&\hphantom{0}9.8\%	&59.8\% \\
HC2	hill-climbing   &+0.1749	&0.8900	$\pm$0.0046	&\hphantom{0}6.8\%	&10.9\%  \\

HC6 hill-climbing   &+0.0781	&0.7516	$\pm$0.0091	&\hphantom{0}4.9\%	&83.5\%  \\

\noalign{\smallskip}\hline
\multicolumn{5}{l}{* Confidence intervals at a level of 0.999. For parameter settings see Table 1.}\\
\end{tabular}

\end{center}
\end{footnotesize}
\end{table}

\begin{figure}
\centering
  \includegraphics[width=0.7\textwidth]{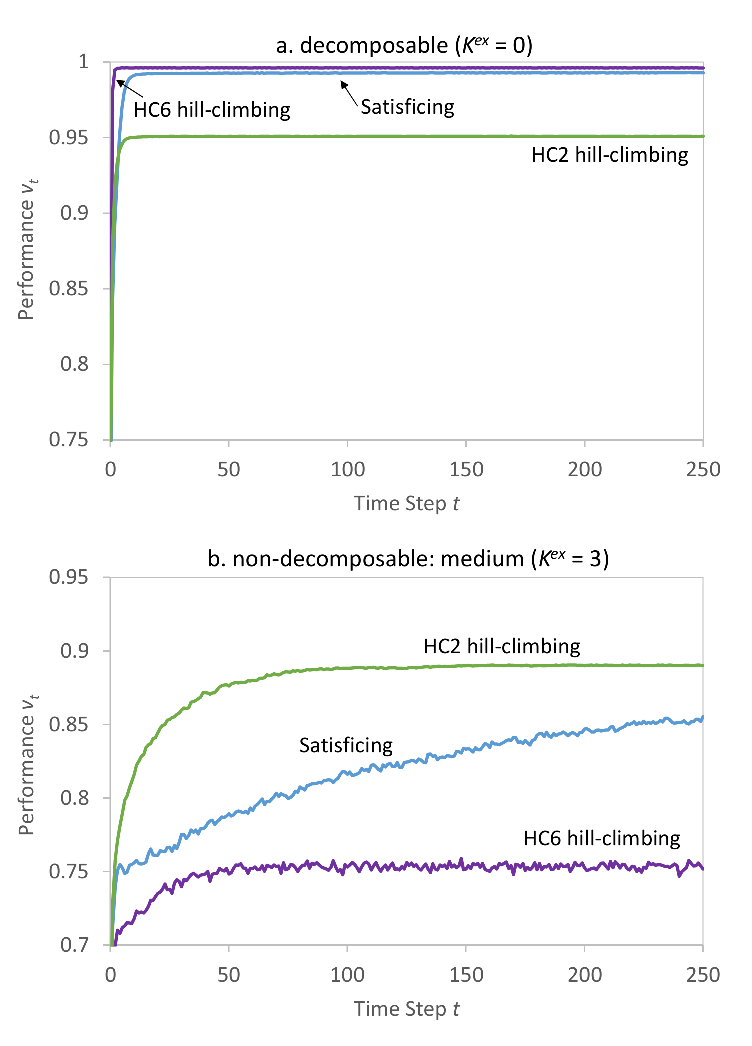}
\caption[ ]{Adaptive walks of the baseline scenarios. Each line represents the average of 2500 simulations. For parameter settings see Table \ref{Parameters}.} \label{baselinewalks} 
\end{figure}

\begin{figure}
\centering
  \includegraphics[width=0.7\textwidth]{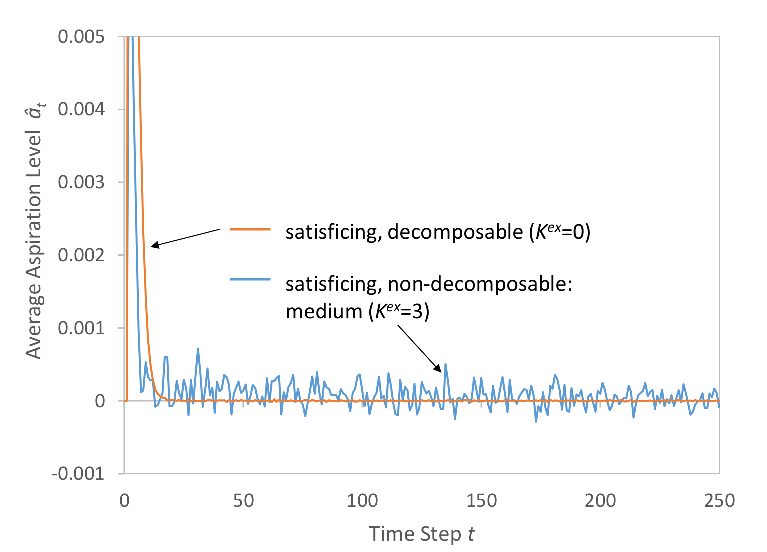}
\caption[ ]{Adaptation of aspiration levels in the baseline scenarios. Each line represents the average of 2500 simulations. For parameter settings see Table \ref{Parameters}.} \label{baselineaspirations} 
\end{figure}

\begin{figure}
\centering
  \hspace*{.05\linewidth}\includegraphics[width=0.7\textwidth]{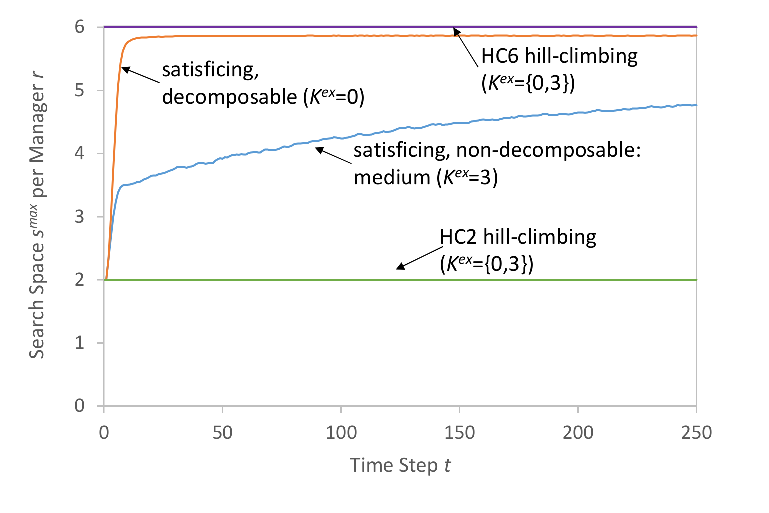}
\caption[ ]{Adaptation of maximum search space per manager in the baseline scenarios. Each line represents the average of 2500 simulations. For parameter settings see Table \ref{Parameters}.} \label{baselinespaces} 
\end{figure}

Figure \ref{baselinewalks} plots the performance levels obtained in the course of adaptive walks over time for each scenario. Figure \ref{baselineaspirations} displays the adaptation of aspiration levels over time for the two satisficing scenarios. Please, recall, in the hill-climbing scenarios, aspiration levels are zero (see Section \ref{hillclimbing}), which is why they are not plotted. Figure \ref{baselinespaces} reports on the adjustment of the search spaces in satisficing search for the decomposable and the non-decomposable structure; the search spaces of the scenarios employing hill-climbing are fixed, as is also indicated in the figure.

The following discussion of results mainly focuses on satisficing search behavior in contrast to the hill-climbing strategies (and less on comparing the hill-climbing modes against each other).

The plots in Figure \ref{baselinewalks} indicate that the performance enhancements obtained via satisficing search behavior are at medium levels compared to the two hill-climbing modes (which, however, perform differently well in the two interaction structures) for both interactions structures under investigation. The results reported in Table \ref{baselineresults} also suggest that satisficing search is at medium levels regarding initial performance enhancements and final performances. For the frequency of global maximum found, satisficing search outperforms both hill-climbing models in the non-decomposable structure. With satisficing, the ratio of periods with altered configurations is at a notably high level compared to the HC2 strategy. In the case of a decomposable structure, it even exceeds the level of the HC6 strategy. For a closer analysis of results, it appears helpful also to consider the adjustment processes of the aspiration levels and maximum search spaces as plotted in Figures \ref{baselineaspirations} and \ref{baselinespaces}, respectively.

\paragraph{{Decomposable interaction structure.}}
\label{sec:DecomposableInteractionStructure}

Each manager faces a partial binary problem in the decomposable interaction structure without any interactions among the managers' problems existing. Hence, the organization’s overall performance maximum could be found by identifying the sub-problems' optimal solutions. Therefore, with a broad search space enabled at the managers' site as with the HC6-strategy, it is not surprising that the adaptive walks quickly reach performance levels close to the maximum of 1. With managers employing satisficing behavior, the performance levels achieved are close to that of the HC6-strategy. Moreover, the maximum number of alternatives per manager increases rather quickly to nearly the high level of 6 as fixed for the HC6 strategy and remains at this high level (see Figure \ref{baselinespaces}). 

The explanation for this is as follows: in the decomposable structure, managers likely find configurations with high or even the maximum performance level for their partial problem. However, from a very high (or maximal) performance level, it becomes more difficult (or impossible) to further increase performance. However, according to the behavioral assumptions underlying the idea of bounded rationality \citep{Simon1955}, the managers are not aware of whether they already have identified the optimal solution. In consequence, since managers experience it difficult to further increase performance, according to the satisficing concept, the search space is increased. It remains at a high level in the -- potentially futile -- attempt to increase performance further. 

The adaption of aspiration levels follows an inverse adjustment: After a high incline in the first periods -- due to high inclines of performance at the beginning -- the aspiration levels decline quickly to a level of zero: with being close to the best configuration (or having it found already), further performance enhancements are unlikely (or even impossible) and, hence, the aspiration levels of decision-making managers, persistently, remain at a level of zero. However, a closer analysis reveals that the aspiration levels \textit{oscillate} closely around zero. This is because the managers in the model are not capable of evaluating options perfectly. Hence, false-positive choices may occur, which then affect the aspiration levels and may turn them to the negative (see Eq. \ref{aspirationadjust}).\footnote{In other words:  When decision-makers can evaluate options perfectly, in the decomposable structure, the aspiration levels do not oscillate around zero. Instead, after an initial incline, they decline to equal zero.}

\paragraph{Non-decomposable interaction structure.}
\label{sec:NonDecomposableInteractionStructure}

In the non-decomposable structure, the link between managers’ sub-problems and the overall decision-problem is more complicated than in the decomposable case for two reasons. First, when searching for superior solutions to their partial decision problems, managers do not necessarily increase overall performance. Hence, maximizing parochial performance and the overall performance of the organization may conflict with each other. The second reason refers to managers’ cognitive limitations regarding their fellow managers’ choices. Due to interactions among sub-problems, manager $r$'s choice for the partial problem $\vec{d}^r$ may affect the performance $P^{q}_{t}(\vec{d^q_t})$ (Eq. \ref{perfmanager}) of another manager $q\neq r$ and vice versa. Since, in the model, the managers notice their fellow managers' choices with one period of delay, this may lead to frequent, time-delayed mutual adjustments in order to keep up with the fellow managers' choices, which again induces mutual adjustments and so forth. These considerations reflect the lower performance levels achieved, the lower frequencies of the global maximum found, and the higher ratios of altered configurations compared to the decomposable structure reported in Table \ref{baselineresults}. These results, in principle, correspond to prior research employing computational models of organizations (\citealp [e.g.,][]{Carley1992,Rivkin2007,Siggelkow2005}). However, the differences across search strategies are remarkable, which is analyzed in more detail in Section \ref{complexity}.

In the satisficing strategy, the adjustments of maximum search spaces and aspiration levels deserve closer inspection. Regarding the adjustment of maximum search spaces in the satisficing strategy (Figure \ref{baselinespaces}), for the non-decomposable structure, we again notice an increase over time -- though up to a lower level of about 5 per manager and with a lower gradient compared to the decomposable structure. This may result from the following effects: as argued above, in non-decomposable structures, it is rather difficult to identify solutions that induce performance enhancements. However, when finding promising options becomes more difficult, with satisficing the maximum search space is increased. At the same time, this may counteract the peril of sticking to local maxima, and the peril of inertia is the more pronounced, the higher the complexity $K$ (or $K^{ex}$) of a decision-problem. \footnote{However, the search space is at a lower level than with satisficing in the decomposable structure. This is because, in the decomposable structure, high levels of performance are found very quickly. With this, further performance enhancements are difficult to achieve, which leads to an extension of the search space close to the upper bound.} 

The adjustment of aspiration levels plotted in Figure \ref{baselineaspirations} shows the inverse development, and aspiration levels decline over time. Contrary to the decomposable case, now the aspiration levels oscillate remarkably around a level of zero. Hence, an interesting question what may cause these oscillations. Like in the decomposable structure, the imperfect evaluations contribute to oscillations of aspiration levels: imperfect ex-ante evaluations may lead to performance declines due to false-positive choices. Accordingly, these ``negative'' experiences are reflected in the adjustment of the aspiration levels. Additionally, in the non-decomposable structure, interactions among sub-problems combined with cognitive limitations regarding the choices of fellow managers further induce oscillating aspiration levels: 
\begin{enumerate}
	\item When making their choices in time step $t$, decision-makers assume that their fellow managers stay with the status quo. This is particularly problematic in case of interactions among decision-problems and may cause ``surprises'' and, in consequence, frequent mutual adjustments (which happens in about 60 percent of periods, see Table \ref{baselineresults}); 
	\item The actual choices of fellow managers are revealed only at the end of period $t$ which causes a time-delay in the aforementioned mutual adjustments to the other managers' choices;	
\end{enumerate}
Hence, due to interactions combined with alterations by fellow managers, performance declines may happen which reduce aspiration levels even below zero.

In sum, intra-organizational complexity in combination with imperfect information in decision-making reasonably causes frequent alterations of configurations $\vec{d}$ and oscillations of aspiration levels in the satisficing strategy. We return to this aspect in Section \ref{complexity}. Taking a more general perspective on the baseline scenarios, one may summarize the findings in the following hypotheses: 

\noindent \textit{(1) Organizations which are resided by decision-makers showing satisficing search behavior and which already have identified configurations providing high levels of performance are likely to employ extensive search and aspiration levels which enforce to (just) maintain the performance. }

\noindent \textit{(2) Intra-organizational complexity combined with cognitive limitations of decision-makers showing satisficing search behavior induces high levels of search activity and oscillating aspiration levels.}

\noindent These hypotheses could be related to organizations’ maturity, and organizational learning in terms of both performance level achieved and organizations' focus on searching for novel solutions. 

\subsection{Sensitivity to Intra-organizational Complexity}\label{complexity}

The next step of analysis considers simulation results for all levels of intra-organizational complexity from $K^{ex}=(0,\ldots5)$. Thereby, we intend to provide more detailed insights into potential differences of satisficing behavior compared to hill-climbing search. For this, Figure \ref {sensitivity} displays -- for the three search strategies under investigation -- (a) the performance level achieved on average of 2500 runs in the last period of observation, (b) the relative frequency of runs in which the global maximum was found in the last period, and (c) the average ratio of periods in which the organizations implement a new solution to their decision problem.  

\begin{figure}
\centering
  \includegraphics[width=0.6\textwidth]{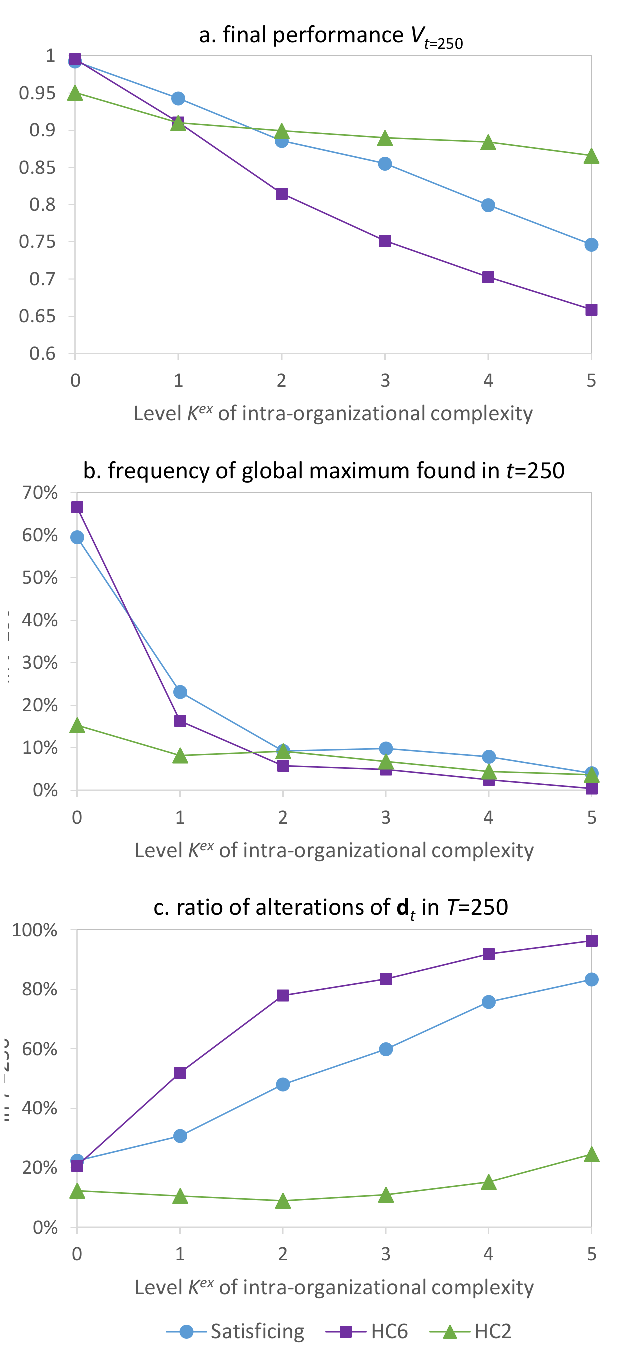}
\caption[ ]{Sensitivity of (a) final performance, (b) frequency of global maximum found and (c) ratio of alterations to intra-organizational complexity. Each mark represents the average of 2500 simulations. For parameter settings see Table \ref{Parameters}.} 
\end{figure}
\label{sensitivity}

The results reveal that, for all search strategies, the final level of performance decreases with increasing intra-organizational complexity, which is broadly in line with prior research \citep[e.g., ][]{Rivkin2007,Levinthal1997}. However, as shown in Figure \ref{sensitivity}.a, the search strategies are differently sensitive to an increase in intra-organizational complexity. The HC2-strategy -- allowing only two alternatives and with just 1-bit changes each -- is comparably robust with about 8.5 percentage points (p.p.) between highest and lowest final performance. In contrast, this difference is about 25 p.p. with sa\-tis\-ficing and 34 p.p. with the HC6-strategy. Hence, these strategies -- allowing for more alternatives considered and longer jumps -- are notably sensitive to intra-organizational complexity in terms of performance declines.

These results might be counter-intuitive since one may expect that search strategies allowing to consider more alternatives and making even longer jumps outperform the HC2-strategy since this strategy is much more ``restrictive'' regarding search space and extent of change. Moreover, concerning satisficing the result is particularly interesting: with this strategy, the decision-makers sequentially discover and ex-ante evaluate alternatives -- and this with increasing Hamming distances starting with two options with 1-bit changed. Hence, intuition may suggest that satisficing should not perform worse but even more successfully than the HC2-strategy. Moreover, it is worth mentioning that the satisficing strategy tends to show higher ratios of locating the optimal solution as Figure \ref{sensitivity}.b suggests.

The more extensive search spaces and longer jumps employed in satisficing -- and likewise with the HC6-strategy -- result in a remarkable increase in alterations as shown in Figure \ref{sensitivity}.c. For example, for high intra-organizational complexity ($K^{ex}=5$), with satisficing in about 83 percent of the periods and with HC6 hill-climbing in almost every period, another solution for the overall decision problem is implemented.

An interesting question is what causes these effects. The explanation may lie in the destabilization of the search when the strategy allows for more alternatives and long jumps as is the case with satisficing and HC6 hill-climbing. In particular, interactions among managers' sub-problems and imperfect information at the managers' site subtly interfere. Each manager $r=(1,\ldots M)$ -- when making its decision in $t$ without knowing what the fellow managers intend to do -- may not only have been surprised by the actual performance $P^r$ achieved in $t-1$. Moreover, the fellow managers' choices in $t-1$ which -- due to intra-organizational interactions -- have affected $r$'s performance in $t-1$ may be another source of surprise for manager $r$. This eventually lets manager $r$ adapt configuration $\vec{d}^r_{t}$ and so forth -- resulting in \textit{frequent time-delayed mutual adjustments}. Hence, search behavior that is more flexible in terms more options and longer jumps makes it more likely that a manager discovers alternatives that (eventually falsely) promise to increase $r$'s performance. In this sense, the flexibility of search may induce some harmful ``hyperactivity’’ of searching when intra-organizational complexity increases. The ratios of alterations increasing in the intra-organizational complexity with satisficing, or HC6 provide support for this conjecture (Figure \ref{sensitivity}.c). 

These considerations may be summarized as follows: Search behavior that is more flexible in terms of considering a higher number of options and longer jumps as captured in satisficing is more prone to destabilizing (``hyperactive'') mutual adjustments than more restrictive forms of search behavior. 

As mentioned before, prior research often employs algorithms like our HC2-strategy to represent local search for superior solutions to organizations’ overall decision problem. In doing so, prior research puts considerable emphasis on complexity, i.e., interactions within the overall decision problem. The sensitivity analysis presented here suggests that satisficing search is remarkably more sensitive to intra-organizational complexity than local search via hill-climbing. This appears particularly relevant since satisficing has received considerable support in behavioral experiments (see Introduction and Section \ref{Simon}), thus, maybe a more realistic computational representation of managerial search behavior than hill-climbing algorithms. 

\section{Conclusion}
At the center of this paper are the questions of representing managerial search behavior in computational models and how the representation may affect models’ results. Prior research questions that hill-climbing algorithms -- predominating in computational organization science -- represent managerial search behavior appropriately. At the same time, there is considerable evidence on the relevance of satisficing behavior in actual human behavior. Against this background, the paper makes two contributions. 

First, the paper introduces an algorithmic representation for backward-looking search according to Simon's concept of satisficing \citep{Simon1955}. The satisficing algorithm may complement other models of managerial search in (agent-based) computational organization science and, in this sense, may contribute to the ongoing discussion on how to model human decision-makers \citep[e.g.,][]{Gode1993,Chen2012,Hommes2006}. 

Second, in an agent-based simulation model of decision-making organizations, the proposed algorithm of satisficing is applied and contrasted to the steepest ascent variant of hill-climbing. Apart from decision-makers’ incomplete knowledge of the solution space, the model captures further aspects of bounded rationality. The simulation experiments suggest that, first, with satisficing for organizations already operating at a high performance level, intense search activities may emerge. Second, oscillating aspiration levels (including accepting performance declines) and potentially destabilizing search activities may occur when intra-organizational complexity is high. Third, a sensitivity analysis reveals that satisficing is considerably more sensitive to intra-organizational complexity in terms of performance declines than hill-climbing algorithms. 

In sum, from a more general perspective, the results suggest that the type of search algorithm the decision-making agents employ (i.e., whether they follow the satisficing concept or a hill-climbing approach) may subtly shape the model’s behavior. These findings may shed some new light on prior modeling efforts building on hill-climbing algorithms, and may even suggest to revisit the respective computational studies in future research efforts\citep[in a similar vein][]{Tracy2017}.

The simulations presented in this paper require relativizing remarks, which also point to future research activities. First of all, it has to be emphasized that the satisficing concept captures some more modeling choices and parameter settings than typically showing up for hill-climbing. This applies particularly to the search sequence and the adjustments of aspiration levels and of the maximum number of options. For example, the simulations presented in this paper assume a ``closest first'' sequence of search and exponential weighting with equal focus on past and presence for the adjustments ``built-in'' in the satisficing concept. Of course, various other types of sequence and adjustments are feasible too. Hence, an obvious further step would be exploring the effects of satisficing on model behavior for a broader parameter space. 

Moreover, the simulation model introduced in this paper captures relatively simple -- for not to say: simplistic – organizations. In particular, the organizational arrangements do not comprise much more than the division of labor (i.e., decomposition into sub-problems and delegation to subordinate managers) and a simple incentive scheme that rewards parochial performance. Hence, an interesting question is how satisficing search behavior shapes results for organizations with more sophisticated institutional arrangements. Of interest may be, for example, how different coordination mechanisms destabilizing effects of satisficing in the case of higher levels of intra-organizational complexity compared to hill-climbing. Studying the satisficing algorithm in models of more sophisticated organizational arrangements will also contribute to linking this representation of managerial search behavior to prior research in computational organization theory.


%
%

\bibliographystyle{apalike}
\bibliography{FWall-Satisficing_ArXiv}

%
%

\end{document}